\documentclass[aps,pra,onecolumn,showpacs]{revtex4}
\usepackage{stmaryrd}
\usepackage{amsfonts}
\usepackage{mathrsfs}
\usepackage{amsmath}
\usepackage{amscd}
\usepackage{graphicx}
\usepackage{booktabs}
\usepackage{leftidx}
\begin{document}
\title{Multiphoton controllable transport between remote resonators}
\author{Wei Qin$^{1,2}$}
\author{Guilu Long$^{3,4}$}\thanks{gllong@mail.tsinghua.edu.cn}
\address{$^1$ School of Physics, Beijing Institute of Technology, Beijing 100081, China\\
$^2$ CEMS, RIKEN, Wako-shi, Saitama 351-0198, Japan\\
$^3$ State Key Laboratory of Low-Dimensional Quantum Physics and Department of Physics, Tsinghua University, Beijing 100084, China\\
$^4$ Tsinghua National Laboratory for Information Science and Technology, Tsinghua University, Beijing 100084, China}

\begin{abstract}
We develop a novel method for multiphoton controllable transport between remote resonators. Specifically, an auxiliary resonator is used to control the coherent long-range coupling of two spatially separated resonators, mediated by a coupled-resonator chain of arbitrary length. In this manner, an arbitrary multiphoton quantum state can be either transmitted through or reflected off the intermediate chain on demand, with very high fidelity. We find, on using a time-independent perturbative treatment, that quantum information leakage of an arbitrary Fock state is limited by two upper bounds, one for the transmitted case and the other for the reflected case. In principle, the two upper bounds can be made arbitrarily small, which is confirmed by numerical simulations.
\end{abstract}
\maketitle
%%%%%%%%%%%%%%%%%%%%%%%%%%%%%%%%%%%%%%%%%%%%%%%%%%%%%%%%%%%%%%%%%%%%%%%%%%%%%%%%%%%%%%%%%
\section{introduction}
\label{se:section1}
The realization of quantum information transport between remote parties is of central importance for any scalable quantum information processing. To this end, one straightforward approach is to use quantum channels to connect these spatially separated parties. Owing to high speed transmission and negligible interaction with the environment, in the form of flying qubits, photons are the natural candidates for long-distance quantum communication \cite{photon1,photon2,photon3,photon4,photon5,photon6,photon7}. Alternatively, for short-distance quantum communication inside a quantum computer, the majority of promising channels rely upon the use of solid-state-based devices, including nuclear spins in nuclear magnetic resonance (NMR) \cite{nuclear1,nuclear2}, electron spins of nitrogen-vacancy (NV) colour centers in diamond \cite{NV1,NV2,NV3,NV4,NV5,NV6}, and flux qubits in superconductors \cite{flux1,flux2,flux3,flux4}. Moreover, coupled-resonator arrays (CRAs), being currently explored in various physical systems such as superconducting transmission line resonators \cite{STLR1,STLR2,STLR3,STLR4,STLR5}, toroidal microresonators \cite{TM1,TM2,TM3,TM4} and plasmonic nanoparticle arrays \cite{PNA}, have been attracted much attention in recent years. A particular advantage of such arrays is the full addressability of individual resonators, which allows each of the resonators to act as a quantum network node \cite{node1,node2,node3}. Indeed, utilizing these CRAs of having the same fundamental hardware to process quantum information can also avoid a quantum interface between the quantum register and the quantum channel.

In addition to offering an effective platform to simulate quantum many-body phenomena such as Mott-superfluid and topological effects \cite{MB1,MB2,MB3,MB4}, the CRAs have been previously considered for controllable transport of photons by making use of the photon-atom scattering \cite{STLR3,CCA_switch1,CCA_switch2,CCA_switch3,CCA_switch4,CCA_switch5}. Despite such substantial developments, prior work on controllable photon transport has typically focused upon either single photons \cite{STLR3,CCA_switch2,single1,single2,single3} or nearby CRAs \cite{multi1_1,multi1_2,multi2_1,multi2_2,multi3}. However, the ability to transport multiphoton quantum states is a key requirement for encoding a high-dimensional Hilbert space, which is applicable, for example, to universal quantum computation \cite{un_qc1,un_qc2,un_qc3}; at the same time, quantum information stored in the multiphoton fields needs to be controllably transported between two distant quantum registers to carry out quantum network operations. For these reasons, developing a quantum channel capable of controllably transporting multiphoton states is thus of both fundamental and practical importance.

In this paper, we propose and analyze a multiphoton controllable transport protocol, where we use an auxiliary resonator coupled to one resonator of a coupled-resonator chain, which serves as a quantum channel to connect two remote resonators. The physical essence underlying our method is that this auxiliary resonator is employed to control the coherent long-range interaction between the two boundary resonators. To be specific, in the case when the auxiliary resonator is absent, the two boundary resonators could be only coupled to a specific eigenmode of the intermediate chain, within the weak-coupling regime. In this case, the time evolution can swap arbitrary bosonic quantum states of the two boundary resonators, yielding an effective photon transport channel (EPTC). On the contrary, when the auxiliary resonator is coupled to the intermediate chain, the specific eigenmode could be split, such that the two boundary resonators are decoupled from the intermediate chain in the large-detuning limit. Photons are therefore reflected back, remaining unchanged. As opposed to prior work, the proposed model is capable of controlling the coherent transport of an arbitrary multiphoton quantum state between two distant resonators over an arbitrarily long range.

%%%%%%%%%%%%%%%%%%%%%%%%%%%%%%%%%%%%%%%%%%
\section{Physical model and effective Hamiltonian}
\label{se:section2}
\begin{figure}[!ht]%[tpb]
\begin{center}
\includegraphics[width=9.5cm,angle=0]{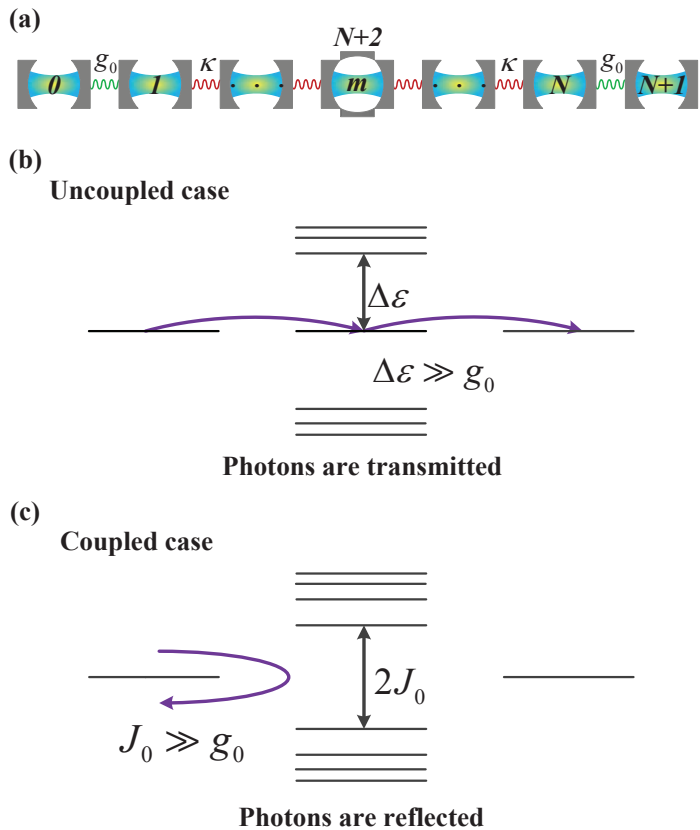}
\caption{(color online) (a) Two remote resonators, labelled $0$ and $N+1$, are coupled to the ends of a coupled-resonator chain of $N$ resonators with $g_{0}$, the coupling strength between the two boundary resonators and the intermediate chain, and $\kappa$, the coupling strength between the intrachain elements. To achieve multiphoton transport, we introduce an auxiliary resonator, labelled $N+2$, to interact with the $m$-th resonator of the intermediate chain,  with strength $J_{0}$. (b) In uncoupled case of $J_{0}=0$, the intermediate chain could, under the assumption that $g_{0}\ll\kappa$, coherently couple the two spatially separated resonators, such that photons are transported between the two boundary resonators after the time evolution. (c) However, when $g_{0}\ll J_{0}\ll \kappa$, the boundary resonators are decoupled from the intermediate chain owing to the large detunings. In this case, photons are reflected back and remains unchanged.}\label{fig1}
\end{center}
\end{figure}

The basic idea of our protocol is schematically illustrated in Fig. \ref{fig1}(a). To begin, we consider a quantum channel consisting of a coupled-resonator chain of $N$ resonators, with a Hamiltonian
\begin{equation}
H_{0}=\kappa\sum_{i=1}^{N-1}\left(c_{i}^{\dag}c_{i+1}+\text{H.c.}\right),
\end{equation}
where $c_{i}$ ($c_{i}^{\dag}$) represents the annihilation (creation) operator acting on the resonator $i$ and obeys a boson commutation relation $\left[c_{i},c_{j}^{\dag}\right]=\delta_{ij}$, and $\kappa$ is the coupling constant between the nearest-neighbor resonators. Two distant resonators, labelled $0$ and $N+1$, are coupled to the ends of the quantum channel, and the interaction Hamiltonian is correspondingly given by
\begin{equation}\label{eq:v1}
V_{1}=g_{0}\left(c_{0}^{\dag}c_{1}+c_{N+1}^{\dag}c_{N}+\text{H.c.}\right),
\end{equation}
where $g_{0}$ is the coupling strength between the two boundary resonators and the intermediate chain. In order to control the multiphoton coherent transport, we introduce an auxiliary control resonator, labelled $N+2$, to interact with the $m$-th resonator of the quantum channel through
\begin{equation}\label{eq:v2}
V_{2}=J_{0}\left(c_{N+2}^{\dag}c_{m}+\text{H.c.}\right),
\end{equation}
with a coupling strength $J_{0}$. Here, we have assumed that all resonators have a common frequency $\omega$, and the system is transformed into a frame rotating at $\omega$. Therefore, the total Hamiltonian governing the system is $H_{T}=H_{0}+V_{1}+V_{2}$.

%%%%%%%%%%%%%%%%%%%%%%%%%%%%%%%%%%%%%%%%%%%%%%%%%%%%%%%%%%%%%%%%%%%%%%%%%%%%%%%%%%%%%%%%%%%%
%%%%%%%%%%%%%%%%%%%%%%%%%%%%%%%%%%%%%%%%%%%%%%%%%%%%%%%%%%%%%%%%%%%%%%%%%%%%%%%%%%%%%%%%%%%%%%

Upon following an orthogonal transformation \cite{ut1,ut2},
\begin{equation}\label{eq:orthog_trans}
f_{k}=\sqrt{\frac{2}{N+1}}\sum_{i=1}^{N}\sin\frac{ik\pi}{N+1}c_{i},
\end{equation}
the Hamiltonian $H_{0}$ is diagonalized to
\begin{equation}
H_{0}=\sum_{k=1}^{N}\varepsilon_{k}f_{k}^{\dag}f_{k},
\end{equation}
where $\varepsilon_{k}=2\kappa\cos\left[k\pi/\left(N+1\right)\right]$. Substituting Eq. (\ref{eq:orthog_trans}) into Eqs. (\ref{eq:v1}) and (\ref{eq:v2}), $V_{1}$ and $V_{2}$ are likewise transformed to
\begin{equation}
V_{1}=\sum_{k=1}^{N}g_{k}\left[c_{0}^{\dag}f_{k}+\left(-1\right)^{k-1}c_{N+1}^{\dag}f_{k}+\text{H.c.}\right],
\end{equation}
and
\begin{equation}
V_{2}=\sum_{k=1}^{N}J_{k}\left(c_{N+2}^{\dag}f_{k}+\text{H.c.}\right),
\end{equation}
respectively. Here, we have defined
\begin{equation}
g_{k}=g_{0}\sqrt{\frac{2}{N+1}}\sin\frac{k\pi}{N+1},
\end{equation}
and
\begin{equation}
J_{k}=J_{0}\sqrt{\frac{2}{N+1}}\sin\frac{mk\pi}{N+1}.
\end{equation}
By restricting our attention to odd $N$, it yields the existence of a single zero-energy mode in the intermediate chain corresponding to $k=z\equiv\left(N+1\right)/2$, such that this mode is in resonance with the two boundary resonators as well as with the auxiliary resonator. Under the assumption that $g_{0}$ and $J_{0}$ are much smaller than $\kappa$, off-resonant couplings to the boundary resonators and to the auxiliary resonator can be neglected owing to $\left\{g_{z},J_{z}\right\}\ll |\varepsilon_{z\pm1}-\varepsilon_{z}|$ \cite{spin0}. As a result, the full evolution dynamics is reduced to an effective model in which only the two boundary resonators, the auxiliary resonator and the zero-energy mode are involved, and accordingly $H_{T}$ is approximated as an effective Hamiltonian
\begin{equation}
H_{\text{eff}}=g_{z}\left[c_{0}^{\dag}f_{z}+\left(-1\right)^{z-1}c_{N+1}^{\dag}f_{z}+\text{H.c.}\right]
+J_{z}\left(c_{N+2}^{\dag}f_{z}+\text{H.c.}\right),
\end{equation}
which can be used to make a multiphoton quantum switch.

%%%%%%%%%%%%%%%%%%%%%%%%%%%%%%%%%%%%%%%%%%%%%%%%%%%%%%%%%%%%%%%%%%%%%%%%%%%%%%%%%%%%%%%%%%%%%%%%%%%%%%%
\section{Multiphoton controllable transport and fidelities}
\label{se:section3}
If the auxiliary resonator is uncoupled to the intermediate chain, $J_{0}=0$, the effective Hamiltonian becomes
\begin{equation}
H_{\text{eff}}=g_{z}\left[c_{0}^{\dag}f_{z}+\left(-1\right)^{z-1}c_{N+1}^{\dag}f_{z}+\text{H.c.}\right].
\end{equation}
In this case, the two boundary resonators are coherently coupled by means of the zero-energy mode. In the Heisenberg picture, a straightforward calculation yields
\begin{equation}
c_{0}^{\dag}\left(t\right)=c_{0}^{\dag}+\frac{1}{2}\left[-1+\cos\left(\sqrt{2}g_{z}t\right)\right]\left[c_{0}^{\dag}+\left(-1\right)^{z-1}c_{N+1}^{\dag}\right]
+\frac{i\sin\left(\sqrt{2}g_{z}t\right)}{\sqrt{2}}f_{z}.\nonumber\\
\end{equation}
Choosing the evolution time $t=\tau\equiv\pi/\sqrt{2}g_{z}$ gives
\begin{equation}\label{eq:switch_on1}
c_{0}^{\dag}\left(\tau\right)=\left(-1\right)^{z}c_{N+1}^{\dag},
\end{equation}
and in a similar manner, we have
\begin{equation}\label{eq:switch_on2}
c_{N+1}^{\dag}\left(\tau\right)=\left(-1\right)^{z}c_{0}^{\dag}.
\end{equation}
Eqs. (\ref{eq:switch_on1}) and (\ref{eq:switch_on2}) exhibit that the evolution for a specific time behaves as a swap operation between the two boundary resonators, as shown in Fig. \ref{fig1}(b).
However, this zero-energy mode could, in the case when the auxiliary resonator is coupled to the intermediate channel, be split into two new modes separated by an energy gap $2J_{z}$. It follows, on ensuring $g_{0}\ll J_{0}$, that the two boundary resonators are significantly detuned from the new modes if $m$ is odd, and thus are decoupled from the intermediate channel. In this case, the time evolution is referred to as an identity operation, leading to
\begin{equation}\label{eq:switch_off1}
c_{0}^{\dag}\left(t\right)=c_{0}^{\dag},
\end{equation}
and
\begin{equation}
c_{N+1}^{\dag}\left(t\right)=c_{N+1}^{\dag}.
\end{equation}

Let us consider the controllable transport of an arbitrary Fock state $|n\rangle$, for example, from the left resonator to the right resonator. The more general treatment of an arbitrary superposition of Fock states are presented in the Appendix. Such a Fock state can be generated using a nonlinear quantum system as an intermediary between a classical radiation field and the resonator \cite{gfock}. We start with an initial state of the total system,
\begin{equation}
|\Phi\left(0\right)\rangle=|n\rangle_{0}|\mathbf{0}\rangle|0\rangle_{N+1}=\frac{\left(c_{0}^{\dag}\right)^{n}}{\sqrt{n!}}|0\rangle_{T},
\end{equation}
where $|\mathbf{0}\rangle=|0\rangle_{1}\cdots|0\rangle_{N}|0\rangle_{N+2}$, and $|0\rangle_{T}=|0\rangle_{0}|\mathbf{0}\rangle|0\rangle_{N+1}$ is the vacuum state of all resonators. Under the time evolution, the total system freely evolves into a finial state,
\begin{equation}\label{eq:state_evolution}
|\Phi\left(t\right)\rangle=\frac{\left[c_{0}^{\dag}\left(-t\right)\right]^{n}}{\sqrt{n!}}|0\rangle_{T}.
\end{equation}
In the uncoupled case when $J_{0}=0$, according to Eq. (\ref{eq:switch_on1}), the finial state for time $\tau$ is $|\Phi\left(\tau\right)\rangle=\left(-1\right)^{nz}|0\rangle_{0}|\mathbf{0}\rangle|n\rangle_{N+1}$,
which means that all photons are simultaneously transported from the left resonator to the right resonator as desired. On the contrary, in the coupled case of $g_{0}\ll J_{0}\ll\kappa$, the finial state becomes $|\Phi\left(t\right)\rangle=|\Phi\left(0\right)\rangle$ according to Eq. (\ref{eq:switch_off1}), thus these photons are reflected back, remaining unchanged.

In order to characterize the quality of our protocol, we employ two fidelities with one transmission fidelity, $F_{t}=\langle n|\rho_{N+1}\left(\tau\right)|n\rangle$, and one reflection fidelity, $F_{r}=\langle n|\rho_{0}\left(\tau\right)|n\rangle$. Here, $\rho_{0}\left(\tau\right)$ and $\rho_{N+1}\left(\tau\right)$ are the reduced density matrices of the resonators $0$ and $N+1$, respectively, at time $t=\tau$.
Together with an $\left(N+1\right)\times\left(N+1\right)$ coupling matrix $A$, the total Hamiltonian can be compactly expressed as
\begin{equation}
H_{T}=\sum_{i,j=0}^{N+2}A_{ij}c_{i}^{\dag}c_{j},
\end{equation}
where $A_{ij}$ is the coupling strength between two resonators $i$ and $j$. Applying the Heisenberg equation of motion, $c_{i}\left(t\right)=i\left[H_{0},c_{i}\left(t\right)\right]$, one finds that
\begin{equation}\label{eq:cm_evolution}
c_{i}\left(t\right)=\sum_{j=0}^{N+2}M_{ij}c_{j},
\end{equation}
with $M=\exp\left(-iAt\right)$ being a unitary evolution matrix. To calculate the two fidelities $F_{t}$ and $F_{r}$, more explicitly, we rewrite $c_{i}\left(t\right)$ of Eq. (\ref{eq:cm_evolution}) as
\begin{equation}\label{eq:operator_evolution}
c_{0}\left(t\right)=M_{0,\mu}c_{\mu}+\sqrt{\delta_{\mu}}c_{\delta_{\mu}},
\end{equation}
where $\delta_{\mu}=1-|M_{0,\mu}|^{2}$ and $\mu=0,\cdots,N+2$. Here, we define a collective mode $c_{\delta_{\mu}}$ as a normalized linear combination of all modes apart from $c_{\mu}$,
resulting in $\left[c_{\mu},c_{\delta_{\mu}}^{\dag}\right]=0$ and $\left[c_{\delta_{\mu}},c_{\delta_{\mu}}^{\dag}\right]=1$. In combination with Eqs. (\ref{eq:state_evolution}) and (\ref{eq:operator_evolution}), we have
\begin{equation}\label{eq:finalstate_fock}
|\Phi\left(t\right)\rangle=\sum_{r=0}^{n}f_{\mu}\left(r,n\right)|r\rangle_{\delta_{\mu}}|n-r\rangle_{\mu},
\end{equation}
where
\begin{equation}\label{eq:fmu}
f_{\mu}\left(r,n\right)=\sqrt{C_{n}^{r}}M_{0,\mu}^{n-r}\delta_{\mu}^{r/2},
\end{equation}
\begin{equation}
C_{n}^{r}=\frac{n!}{r!\left(n-r\right)!},
\end{equation}
and $|r\rangle_{\delta_{\mu}}$ is a Fock state of the collective mode $c_{\delta_{\mu}}$. The corresponding density matrix is
\begin{eqnarray}
&\rho\left(t\right)&=|\Phi\left(t\right)\rangle\langle\Phi\left(t\right)|\nonumber\\
&&=\sum_{r,r'=0}^{n}f_{\mu}\left(r,n\right)f_{\mu}^{*}\left(r',n\right)|n-r\rangle_{\mu}\langle n-r'|\otimes|r\rangle_{\delta_{\mu}}\langle r'|,
\end{eqnarray}
and then, by tracing out the variables of the collective mode, the reduced density matrix of the resonator $\mu$ is calculated as
\begin{equation}
\rho_{\mu}\left(t\right)=\sum_{r=0}^{n}|f_{\mu}\left(r,n\right)|^{2}|n-r\rangle_{\mu}\langle n-r|.
\end{equation}
Thus, the transmission fidelity and the reflection fidelity are respectively given by
\begin{equation}
F_{t}=|f_{N+1}\left(0,n\right)|^{2}=|M_{0,N+1}|^{2n},
\end{equation}
and
\begin{equation}
F_{r}=|f_{0}\left(0,n\right)|^{2}=|M_{0,0}|^{2n}.
\end{equation}

\begin{figure}[!ht]%[tpb]
\begin{center}
\includegraphics[width=8.5cm,angle=0]{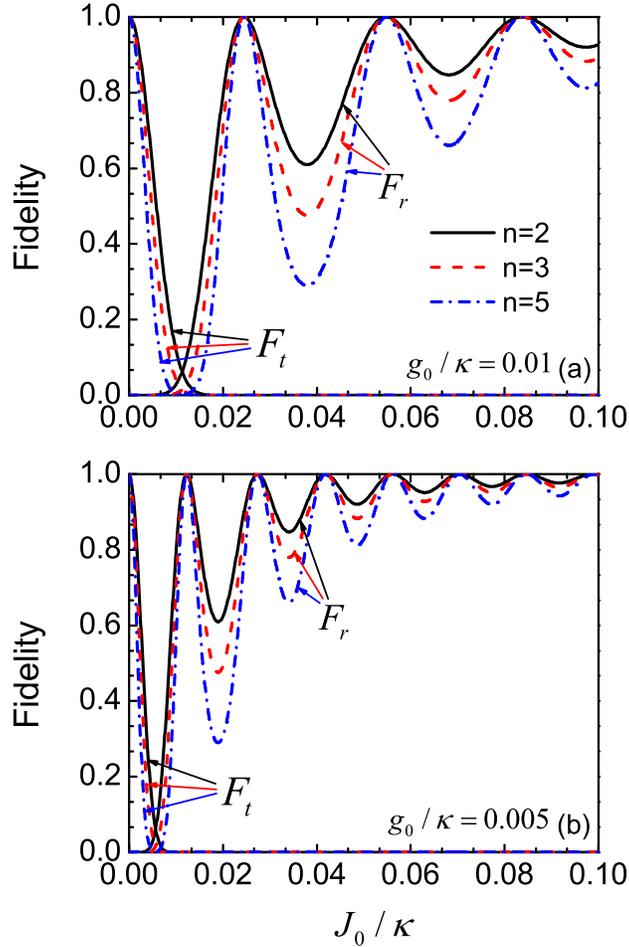}
\caption{(color online) Transmission fidelity $F_{t}$ and reflection fidelity $F_{r}$ as a function of $J_{0}/\kappa$ with either (a) $g_{0}/\kappa=0.01$ or (b) $0.005$ for $n=2,3$ and $5$, corresponding to solid black lines, dashed red lines and dot-dashed blue lines, respectively. Here, the evolution time $t=\tau$ and we choose $N=7$, $m=3$.}\label{fig2}
\end{center}
\end{figure}

In Fig. \ref{fig2} we plot the transmission fidelity $F_{t}$ versus $J_{0}/\kappa$ with  $g_{0}/\kappa=0.01$ [see Fig. \ref{fig2}(a)], or $0.005$ [see Fig. \ref{fig2}(b)], as well as the reflection fidelity $F_{r}$. In the two cases, we assume that a Fock state $|n\rangle$ is initialized into the left boundary resonator, and further take $n=2,3$ and $5$ as three special examples to simulate numerically. It is shown that as $J_{0}=0$, we obtain $F_{t}\simeq1$ while $F_{r}\simeq0$, implying that photons are transmitted from the left resonator to the right resonator. On the contrary, in the regime $g_{0}\ll J_{0}\ll \kappa$, photons are reflected back to the left resonator, according to $F_{t}\simeq0$ whereas $F_{r}\simeq1$. Hence, a Fock state could, by tuning the coupling strength $J_{0}$, be transmitted forward or reflected backward at will.

%%%%%%%%%%%%%%%%%%%%%%%%%%%%%%%%%%%%%%%%%%%%%%%%%%%%%%%%%%%%%%%%%%%%%%%%%%%%%%%%%%%%%%%%%%%%
%%%%%%%%%%%%%%%%%%%%%%%%%%%%%%%%%%%%%%%%%%%%%%%%%%%%%%%%%%%%%%%%%%%%%%%%%%%%%%%%%%%%%%%%%%%%%%

%%%%%%%%%%%%%%%%%%%%%%%%%%%%%%%%%%%%%%%%%%%%%%%%%%%%%%%%%%%%%%%%%%%%%%%%%%%%%%%%%%%%%%%%%%%%%%%%%%%%%%%%%%%%%%%%%%%%%%%%%%%%%%%%%%%%%%%%%%%%%%%%%%%%%%%%%%%%%%%%%%%%%%%%%%%%%%%%%%%%%%%%%%%%%%%%
%%%%%%%%%%%%%%%%%%%%%%%%%%%%%%%%%%%%%%%%%%%%%%%%%%%%%%%%%%%%%%%%%%%%%%%%%%%%%%%%%%%%%%%%%%%%%%%%%%%%%%%%%%%%%%%%%%%%%%%%%%%%%%%%%%%%%%%%%%%%%%%%%%%%%%%%%%%%%%%%%%%%%%%%%%%%%%%%%%%%%%%%%%%%%%%%
\section{Quantum information leakage}
\label{se:section4}
Having explicitly demonstrated the controllable eigenmode-mediated transport of multiphoton information, we now calculate the quantum information leakage by making use of perturbation theory. We begin by considering the evolution matrix of Eq. (\ref{eq:cm_evolution}). In fact, the coupling matrix $A$ is identical to the total Hamiltonian confined in a single-excitation subspace,
\begin{equation}
H_{S}=g_{0}\left(|0\rangle\langle1|+|N+1\rangle\langle N|+\text{H.c.}\right)
+\kappa\sum_{j=1}^{N-1}\left(|j\rangle\langle j+1|+\text{H.c.}\right)
+J_{0}\left(|N+2\rangle\langle m|+\text{H.c.}\right),
\end{equation}
where $|\mu\rangle=c_{\mu}^{\dag}|0\rangle_{T}$ $(\mu=0,\cdots,N+2)$ represents the $\mu$-th basis of this single-excitation subspace. As a consequence, the evolution matrix $M$
is identical to $U\left(t\right)=e^{-iH_{S}t}$, so that $M_{\mu,\mu'}=\langle \mu|U\left(t\right)|\mu'\rangle$. By introducing an orthogonal transformation,
\begin{equation}
|k\rangle=\sqrt{\frac{2}{N+1}}\sum_{j=1}^{N}\sin\frac{jk\pi}{N+1}|j\rangle,
\end{equation}
the Hamiltonian $H_{S}$ is transformed to
\begin{equation}
H_{S}=\sum_{k=1}^{N}g_{k}\left[|0\rangle\langle k|+\left(-1\right)^{k-1}|N+1\rangle\langle k|+\text{H.c.}\right]
+\sum_{k=1}^{N}\varepsilon_{k}|k\rangle\langle k|+\sum_{k=1}^{N}J_{k}\left(|N+2\rangle\langle k|+\text{H.c.}\right).
\end{equation}
Based upon such a Hamiltonian, we can obtain the leakage of quantum information for the uncoupled case of $J_{0}=0$, and the coupled case of $g_{0}\ll J_{0}\ll\kappa$, respectively, as we will show in the following.

%%%%%%%%%%%%%%%%%%%%%%%%%%%%%%%%%%%%%%%%%%%%%%%%%%%%%%%%%%%%%%%%%%%%%%%%%%%%%%%%%%%%%%%%%%%%
\begin{figure}[!ht]%[tpb]
\begin{center}
\includegraphics[width=8.5cm,angle=0]{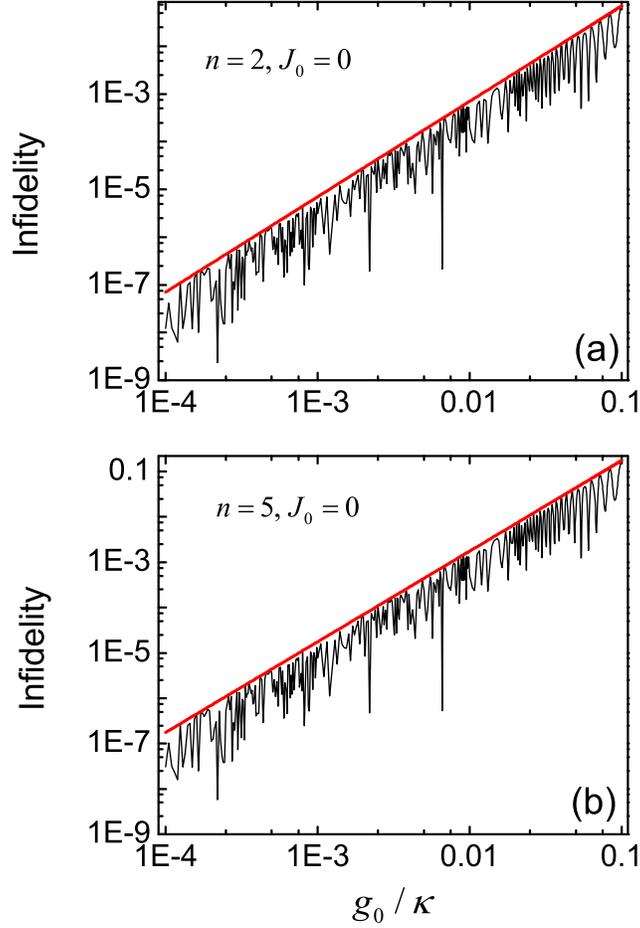}
\caption{(color online) Numerical results of the transmission infidelity, $\sigma_{t}=1-F_{t}$, as a function of $g_{0}/\kappa$ with either (a) $n=2$ or (b) $n=5$ in the uncoupled case of $J_{0}=0$. The analytic upper bounds are represented by the bold red lines. Here, the evolution time $t=\tau$ and we choose $N=7$, $m=3$.}\label{fig3}
\end{center}
\end{figure}
%%%%%%%%%%%%%%%%%%%%%%%%%%%%%%%%%%%%%%%%%%%%%%%%%%%%%%%%%%%%%%%%%%%%%%%%%%%%%%%%%%%%%%%%%%%%%%

Let us first focus upon the former case, where the total system can be thought of as an EPTC being perturbatively coupled to a fictitious bosonic environment in the limit $g_{0}\ll\kappa$. The corresponding Hamiltonian could be divided into three parts, $H_{S}=H_{\text{E}}+H_{z}+V_{z}$. $H_{\text{E}}$ features the Hamiltonian for the EPTC:
\begin{equation}
H_{\text{E}}=g_{z}\left[|e_{1}\rangle\langle e_{2}|+\left(-1\right)^{z-1}|e_{3}\rangle\langle e_{2}|+\text{H.c.}\right],
\end{equation}
where, for convenient, we have used $\left\{|e_{1}\rangle,|e_{2}\rangle,|e_{3}\rangle\right\}$ to replace $\left\{|0\rangle,|z\rangle,|N+1\rangle\right\}$.
The environment is determined by
\begin{equation}
H_{z}=\sum_{k\neq z}\varepsilon_{k}|k\rangle\langle k|
\end{equation}
of having $2N$ bosonic modes, and $\varepsilon_{k}=\varepsilon_{-k}$. The part $V_{z}$ modelling the interaction between them is
\begin{equation}
V_{z}=\sum_{k\neq z}g_{k}\left[|e_{1}\rangle\langle k|+\left(-1\right)^{k-1}|e_{3}\rangle\langle k|+\text{H.c.}\right].
\end{equation}
Assuming that the EPTC Hamiltonian can be diagonalized through a unitary transformation $T$, it results in $\sum_{i,j=1}^{3}T^{\dag}_{q',i}T_{j,q}\langle e_{i}|H_{\text{E}}|e_{j}\rangle=\lambda_{q}\delta_{q',q}$, such that
\begin{equation}
H_{\text{E}}=\sum_{q=1}^{3}\lambda_{q}|q\rangle\langle q|,
\end{equation}
with $|q\rangle=\sum_{i=1}^{3}T_{iq}|e_{i}\rangle$. Similarly,
\begin{equation}
V_{z}=\sum_{k\neq z}\sum_{q=1}^{3}\left(G_{kq}|q\rangle\langle k|+\text{H.c.}\right),
\end{equation}
where
\begin{equation}
G_{kq}=g_{k}\left[T_{q,1}^{\dag}+\left(-1\right)^{k-1}T_{q,3}^{\dag}\right].
\end{equation}
By performing a first-order perturbative treatment, $V_{z}$ has no effects on the eigenenergies $\lambda_{q}$ and $\varepsilon_{k}$; nevertheless, the eigenstates $|q\rangle$ and $|k\rangle$ are modified by
\begin{equation}\label{eq:modefied_q}
|\tilde{q}\rangle\simeq|q\rangle-\sum_{k\neq z}\frac{G_{kq}^{*}}{\varepsilon_{k}}|k\rangle,
\end{equation}
and
\begin{equation}\label{eq:modefied_k}
|\tilde{k}\rangle\simeq|k\rangle+\sum_{q=1}^{3}\frac{G_{kq}}{\varepsilon_{k}}|q\rangle,
\end{equation}
respectively. Combining Eqs. (\ref{eq:modefied_q}) and (\ref{eq:modefied_k}), we find, up to second order,
\begin{equation}
|q\rangle\simeq\left(1-\sum_{k\neq z}\frac{|G_{kq}|^{2}}{\varepsilon_{k}^{2}}\right)|\tilde{q}\rangle
+\sum_{k\neq z}\frac{G_{kq}^{*}}{\varepsilon_{k}}|\tilde{k}\rangle-\sum_{k\neq z}\sum_{q'\neq q}\frac{G_{kq}^{*}G_{kq'}}{\varepsilon_{k}^{2}}|q'\rangle,
\end{equation}
and hence, after an iteration, $|q\rangle$ takes the form of
\begin{equation}
|q\rangle\simeq|\tilde{q}\rangle+\sum_{k\neq z}\frac{G_{kq}^{*}}{\varepsilon_{k}}|\tilde{k}\rangle-\sum_{k\neq z}\sum_{q'=1}^{3}\frac{G_{kq}^{*}G_{kq'}}{\varepsilon_{k}^{2}}|\tilde{q'}\rangle.
\end{equation}
Since
\begin{eqnarray}
&&\langle \tilde{q'}|\tilde{q}\rangle=\delta_{q',q}+\sum_{k\neq z}\frac{G_{kq}^{*}G_{kq'}}{\varepsilon_{k}^{2}},\nonumber\\
&&\langle \tilde{k'}|\tilde{k}\rangle=\delta_{k'k}+\sum_{q}\frac{G_{k'q}^{*}G_{kq}}{\varepsilon_{k'}\varepsilon_{k}},\nonumber\\
&&\langle \tilde{q}|\tilde{k}\rangle=0,
\end{eqnarray}
the matrix elements of $U\left(t\right)$ in the energy space can be evaluated as
\begin{equation}
\langle q'|U\left(t\right)|q\rangle\simeq e^{-i\lambda_{q}t}\delta_{q',q}-\sum_{k\neq z}\frac{G_{kq}^{*}G_{kq'}}{\varepsilon_{k}^{2}}e^{-i\lambda_{q'}t}+\sum_{k\neq z}\frac{G_{kq'}G_{kq}^{*}}{\varepsilon_{k}^{2}}e^{-i\varepsilon_{k}t}.
\end{equation}
Subsequently, after inversion back to the space spanned by $\left\{|e_{i}\rangle\right\}$ ($i=1,2,3$), we arrive at
\begin{multline}
\langle e_{1}|U\left(t\right)|e_{3}\rangle\simeq\langle e_{1}|e^{-iH_{\text{E}}t}|e_{3}\rangle\\+\sum_{k\neq z}\left(-1\right)^{k-1}\frac{g_{k}^{2}}{\varepsilon_{k}^{2}}e^{-i\varepsilon_{k}t}
-\sum_{k\neq z}\frac{g_{k}^{2}}{\varepsilon_{k}^{2}}\left[\left(-1\right)^{k-1}\langle e_{1}|e^{-iH_{\text{E}}t}|e_{1}\rangle+\langle e_{1}|e^{-iH_{\text{E}}t}|e_{3}\rangle\right].
\end{multline}
Combining the evolution under $H_{E}$ leads to $\langle e_{1}|e^{-iH_{\text{E}}\tau}|e_{3}\rangle=\left(-1\right)^{z}$ and $\langle e_{1}|e^{-iH_{\text{E}}\tau}|e_{1}\rangle=0$, and further we find
\begin{equation}
M_{0,N+1}\simeq\left(-1\right)^{z}\left(1-2\Delta_{t}\right),
\end{equation}
for the evolution time $\tau$. Here, we have defined
\begin{equation}\label{eq:Deltat}
\Delta_{t}=\sum_{k<z}\frac{g_{k}^{2}}{\varepsilon_{k}^{2}}\left[1-\left(-1\right)^{k+z-1}\cos\left(\varepsilon_{k}\tau\right)\right].
\end{equation}
The transmission fidelity is therefore modified by
\begin{eqnarray}
F_{t}\simeq(1-2\Delta_{t})^{2n}\simeq1-4n\Delta_{t}.
\end{eqnarray}
The leakage of quantum information can be quantified using a transmission infidelity, $\sigma_{t}=1-F_{t}\simeq4n\Delta_{t}$, and thus has an upper bound,
\begin{equation}
\sigma_{t}\leq8n\sum_{k<z}\frac{g_{k}^{2}}{\varepsilon_{k}^{2}}.
\end{equation}

%%%%%%%%%%%%%%%%%%%%%%%%%%%%%%%%%%%%%%%%%%%%%%%%%%%%%%%%%%%%%%%%%%%%%%%%%%%%%%%%%%%%%%%%%%%%
\begin{figure}[!ht]%[tpb]
\begin{center}
\includegraphics[width=8.5cm,angle=0]{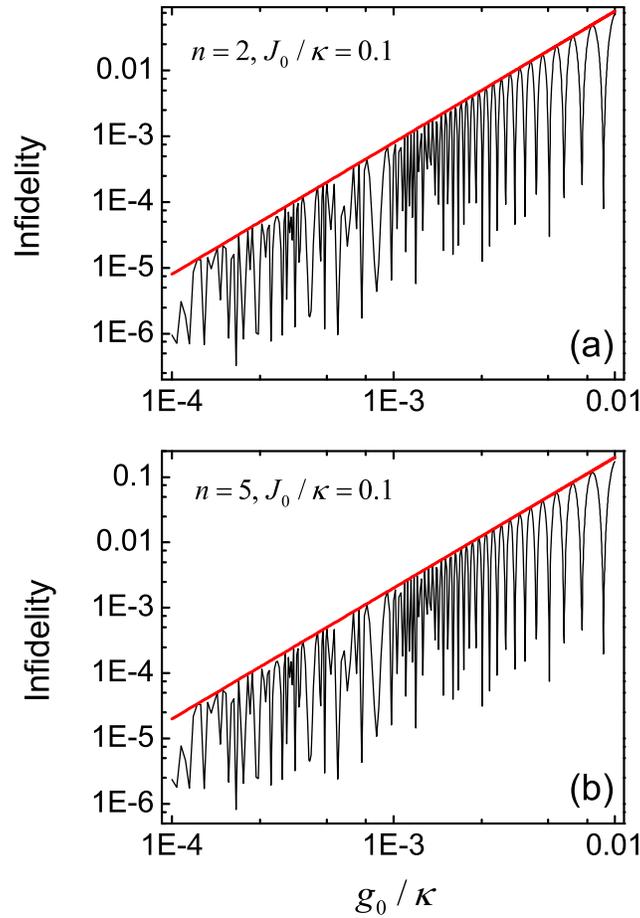}
\caption{(color online) Numerical results of the reflection infidelity, $\sigma_{r}=1-F_{r}$, as a function of $g_{0}/\kappa$ with either (a) $n=2$ or (b) $n=5$ in the coupled case of $J_{0}/\kappa=0.1$. The analytic upper bounds are represented by the bold red lines. Here, the evolution time $t=\tau$ and we choose $N=7$, $m=3$.}\label{fig4}
\end{center}
\end{figure}
%%%%%%%%%%%%%%%%%%%%%%%%%%%%%%%%%%%%%%%%%%%%%%%%%%%%%%%%%%%%%%%%%%%%%%%%%%%%%%%%%%%%%%%%%%%%%%

We now consider the coupled case. In this case, the zero-energy mode of the intermediate chain is the only state coupled to the auxiliary resonator due to $J_{0}\ll\kappa$, which induces two new modes as previously mentioned. The two boundary resonators are coupled to such new modes in addition to the fictitious environment; however, this coupling to the environment can be neglected so long as $g_{0}\ll J_{0}$. The Hamiltonian is therefore reduced to
\begin{equation}\label{eq:H_Origin_coupl}
H_{S}=g_{z}\left[|e_{1}\rangle\langle e_{2}|+\left(-1\right)^{z-1}|e_{3}\rangle\langle e_{2}|+\text{H.c.}\right]
+J_{z}\left(|e_{4}\rangle\langle e_{2}|+\text{H.c.}\right).
\end{equation}
Here, we have used $\left\{|e_{1}\rangle,|e_{2}\rangle,|e_{3}\rangle,|e_{4}\rangle\right\}$ to replace $\left\{|0\rangle,|z\rangle,|N+1\rangle,|N+2\rangle\right\}$ for convenient.
Using $|\gamma_{1}\rangle=\left(|e_{2}\rangle+|e_{4}\rangle\right)/2$, $|\gamma_{2}\rangle=\left(|e_{2}\rangle-|e_{4}\rangle\right)/2$ and $|\pm\rangle=\left(|e_{1}\rangle\pm|e_{3}\rangle\right)/\sqrt{2}$, the Hamiltonian $H_{S}$ of Eq. (\ref{eq:H_Origin_coupl}) is brought to
\begin{equation}
H_{S}=\frac{g_{z}}{2}\sum_{k=1,2}\bigg[\big(|+\rangle+|-\rangle\big)\langle\gamma_{k}|+\left(-1\right)^{z-1}\big(|+\rangle
-|-\rangle\big)\langle\gamma_{k}|+\text{H.c.}\bigg]+\sum_{k=1,2}J_{k}|\gamma_{k}\rangle\langle\gamma_{k}|,
\end{equation}
with $J_{1}=-J_{2}=J_{z}$. When $\left(-1\right)^{z-1}=1$, we get
\begin{equation}\label{eq:couped_H}
H_{S}=g_{z}\sum_{k=1,2}\left(|+\rangle\langle\gamma_{k}|+\text{H.c.}\right)+\sum_{k=1,2}J_{k}|\gamma_{k}\rangle\langle\gamma_{k}|,
\end{equation}
from which we can follow the same recipe as described in the uncoupled case to obtain
\begin{equation}\label{eq:m00}
M_{0,0}\simeq1-2\Delta_{r},
\end{equation}
where
\begin{equation}
\Delta_{r}=\frac{g_{z}^{2}}{2J_{z}^{2}}\left[1-\cos\left(J_{z}t\right)\right].
\end{equation}
For $\left(-1\right)^{z-1}=-1$, it has the same result as mentioned in Eq. (\ref{eq:m00}). In direct analogy to the uncoupled case, the reflection fidelity is modified by
\begin{equation}
F_{r}\simeq\left(1-2\Delta_{r}\right)^{2n}\simeq1-4n\Delta_{r}.
\end{equation}
Correspondingly, the reflection infidelity is $\sigma_{r}=1-F_{r}\simeq4n\Delta_{r}$, together with an upper bound,
\begin{equation}
\sigma_{r}\leq4n\frac{g_{z}^{2}}{J_{z}^{2}}.
\end{equation}

To confirm our calculation of quantum information leaking into the off-resonant couplings \cite{numcals}, we compare numerical results of the transmission infidelity to the analytical upper bound as depicted in Fig. \ref{fig3}, as well as the reflection infidelity in Fig. \ref{fig4}. It is found that this upper bound is in excellent agreement with the numerical results. In addition, the leakage of quantum information decreases with deceasing $g_{0}/\kappa$, so that this leakage can, in principle, be made arbitrarily small.

%%%%%%%%%%%%%%%%%%%%%%%%%%%%%%%%%%%%%%%%%%

\section{Conclusions}
\label{se:section5}
In this paper, we have proposed a new approach for multiphoton controllable transport between two remote resonators being coupled to the ends of a coupled-resonator chain of arbitrary length.
This manner essentially enables a coherent long-range interaction between the two spatially separated resonators, in this case, the pure Hamiltonian evolution for a specific time is referred to as a swap operation of
the two boundary resonators. As a result, an arbitrary multiphoton quantum state can be transported with quantum information leakage arbitrarily close to zero. However, if
an auxiliary resonator is harnessed to coupled to one cavity of the intermediate chain, this coherent long-range interaction will be eliminated, so the two boundary resonators
are decoupled from the intermediate chain, yielding that the time evolution functions as an identity operation. Thus, an arbitrary multiphoton quantum state can be reflected
back with quantum information leakage also arbitrarily close to zero. Our approach potentially allows for realizing controllable transport of an arbitrary dimensional
quantum state or even a coherent state [see the Appendix], and can also be directly generalized to quantum networks consisting of at least three quantum registers. In fact, although we have discussed the specific case of a coupled-resonator chain, such a description is consistent for quantum coupled spin systems of having been widely studied \cite{ut1,ut2,spin0,spin1,spin2,spin3,spin4,spin5}. Hence, the present approach could be applied to the realization of scalable quantum devices, for example, quantum switches.

%%%%%%%%%%%%%%%%%%%%%%%%%%%%%%%%%%%%%%%%%%%%%%%%%%%%%%%%%%%%%%%%%%%%%%%%%%%%%%%%%%%%%%%%%%%%%%%%%%%%%%%%%%%%%%%%%%%%%%%%%%%%%%%%%%%%%%%%%%%%%%%%%%%%%%%%%%%%%%%%%%%%%%%%%%%%%%%%%
\section{acknowledgments}
This work was supported by the National Natural Science Foundation of China under Grant No. 11175094 and 91221205, the National Basic Research Program of China under Grant No. 2015CB921002, and the Basic Research Fund of the Beijing Institute of Technology under Grant No. 20141842005.

%% optional
\setcounter{figure}{0}
\renewcommand{\thefigure}{A\arabic{figure}}
\setcounter{equation}{0}
\renewcommand{\theequation}{A\arabic{equation}}
\appendix %%MDPI internal note: new layout%%

\section*{\noindent Appendix}\vspace{6pt} %%MDPI internal note: new layout%%
In this appendix, we will discuss the controllable transport of an arbitrary $d$-dimensional multiphoton state $|\psi\rangle$, which is a linear superposition of Fock states, $|\psi\rangle=\sum_{n=0}^{d-1}\alpha_{n}|n\rangle$. When preparing this state in the left resonator, the initial state of the total system then is
%%%%%%%%%%%%%%%%%%%%%%%%%%%%%%%%%%%%%%%%%%%%%%%%%%%%%%%%%%%%%%%%%%%%%%%%%%%%%%%%%%%%%%%%%%%
\begin{figure}[!ht]%[tpb]
\begin{center}
\includegraphics[width=8.5cm,angle=0]{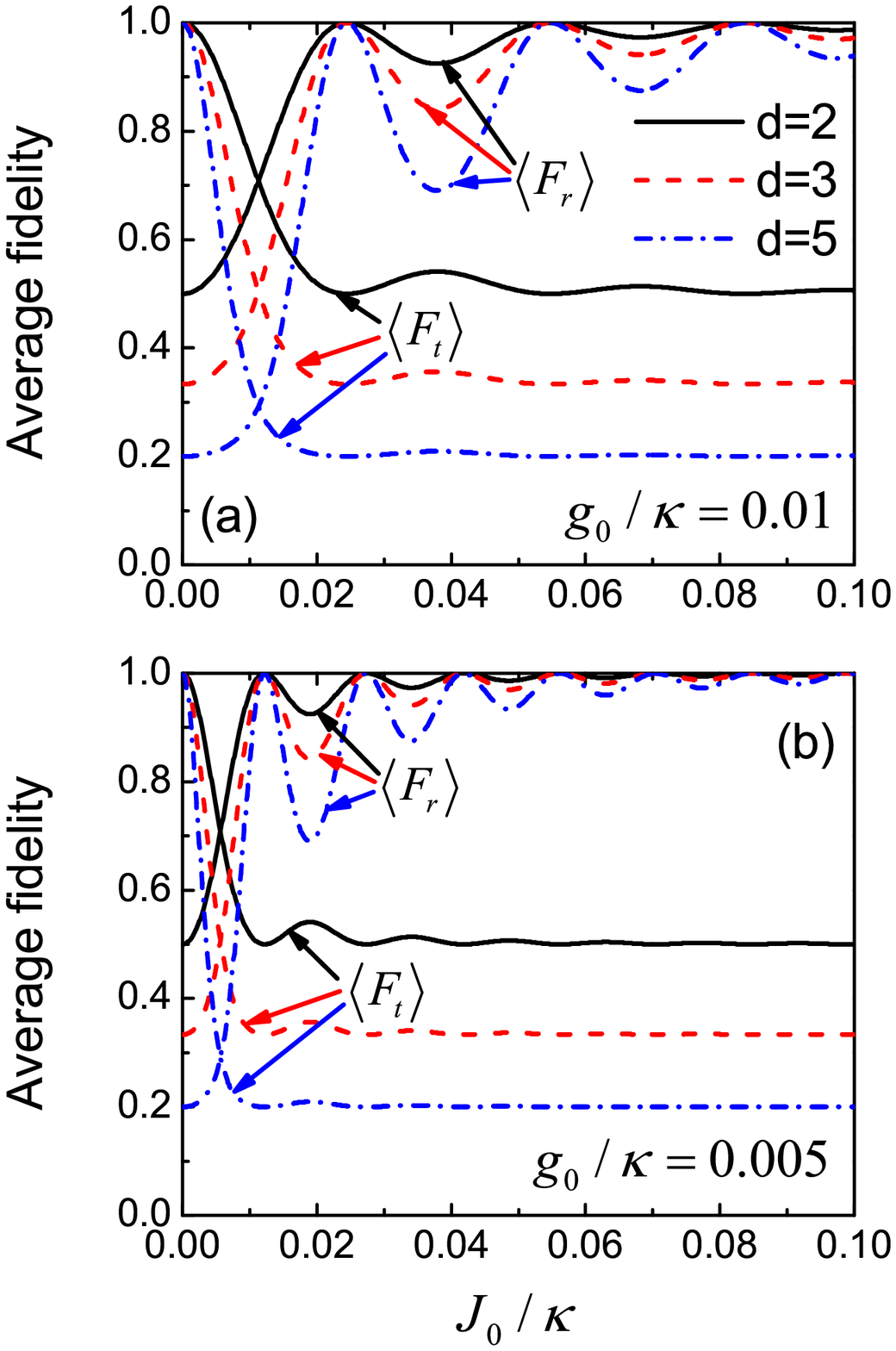}
\caption{(color online) Average transmission fidelity $\langle F_{t}\rangle$ and average reflection fidelity $\langle F_{r}\rangle$ as a function of $J_{0}/\kappa$ with either (a) $g_{0}/\kappa=0.01$ or (b) $0.005$ for $d=2,3$ and $5$, corresponding to solid black lines, dashed red lines and dot-dashed blue lines, respectively. Here, the evolution time $t=\tau$ and we choose $N=7$, $m=3$.}\label{figa1}
\end{center}
\end{figure}
%%%%%%%%%%%%%%%%%%%%%%%%%%%%%%%%%%%%%%%%%%%%%%%%%%%%%%%%%%%%%%%%%%%%%%%%%%%%%%%%%%%%%%%%%%%%%%
\begin{equation}
|\Phi\left(0\right)\rangle=|\psi\rangle|\mathbf{0}\rangle|0\rangle_{N+1}=\sum_{n=0}^{d-1}\frac{\alpha_{n}\left(c_{0}^{\dag}\right)^{n}}{\sqrt{n!}}|0\rangle_{T}.
\end{equation}
Under the total Hamiltonian $H_{T}$, according to Eq. (\ref{eq:finalstate_fock}), the time evolution results in
\begin{equation}
|\Phi\left(t\right)\rangle=\sum_{n=0}^{d-1}\sum_{r=0}^{n}\alpha_{n}f_{\mu}\left(r,n\right)|r\rangle_{\delta_{\mu}}|n-r\rangle_{\mu},
\end{equation}
so the corresponding density matrix could be expressed as
\begin{equation}
\rho\left(t\right)=\sum_{n,n'=0}^{d-1}\sum_{r=0}^{n}\sum_{r'=0}^{n'}\alpha_{n}\alpha^{*}_{n'}f_{\mu}\left(r,n\right)f_{\mu}^{*}\left(r',n'\right)|n-r\rangle_{\mu}\langle n'-r'|
\otimes |r\rangle_{\delta_{\mu}}\langle r'|.
\end{equation}
Consequently, by tracing out the variables of the collective mode, the reduced density matrix of the resonator $\mu$ is evaluated as
%%%%%%%%%%%%%%%%%%%%%%%%%%%%%%%%%%%%%%%%%%%%%%%%%%%%%%%%%%%%%%%%%%%%%%%%%%%%%%%%%%%%%%%%%%%
\begin{figure}[!ht]%[tpb]
\begin{center}
\includegraphics[width=8.5cm,angle=0]{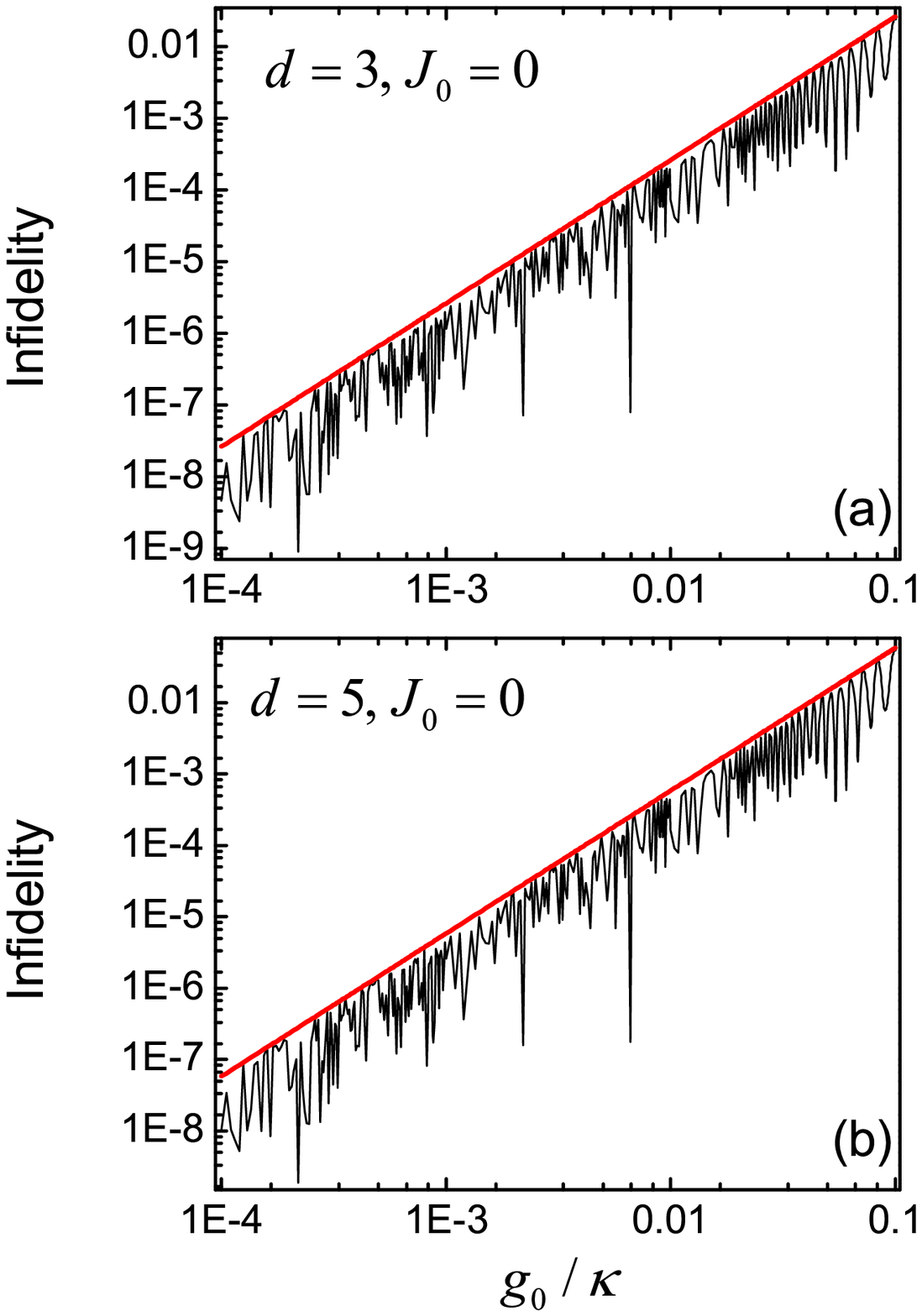}
\caption{(color online) Numerical results of the average transmission infidelity, $\langle \sigma_{t}\rangle=1-\langle F_{t}\rangle$, as a function of $g_{0}/\kappa$ with either (a) $d=3$ or (b) $d=5$ in the uncoupled case of $J_{0}=0$. The analytic upper bounds are represented by the bold red lines. Here, the evolution time $t=\tau$ and we choose $N=7$, $m=3$.}\label{figa2}
\end{center}
\end{figure}
%%%%%%%%%%%%%%%%%%%%%%%%%%%%%%%%%%%%%%%%%%%%%%%%%%%%%%%%%%%%%%%%%%%%%%%%%%%%%%%%%%%%%%%%%%%%%%
\begin{equation}
\rho_{\mu}\left(t\right)=\sum_{n,n'=0}^{d-1}\sum_{r=0}^{\text{min}\{n,n'\}}\alpha_{n}\alpha_{n'}^{*}f_{\mu}\left(r,n\right)f_{\mu}^{*}\left(r,n'\right)|n-r\rangle_{\mu}\langle n'-r\rangle.
\end{equation}
After a straightforward computation, the transmission and reflection fidelities are given, respectively, by
\begin{equation}
F_{t}\equiv\langle \psi|P^{\dag}\rho_{N+1}\left(\tau\right)P|\psi\rangle
=\sum_{n,n'=0}^{d-1}\sum_{r=0}^{\text{min}\{n,n'\}}\left(-1\right)^{\left(n+n'\right)z}\alpha_{n}\alpha_{n'-r}\alpha_{n'}^{*}\alpha_{n-r}^{*}f_{N+1}
\left(r,n\right)f_{N+1}^{*}\left(r,n'\right),
\end{equation}
along with $P=\exp\{iz\pi c_{N+1}^{\dag}c_{N+1}\}$ to ensure the right phase evolution, and
\begin{equation}
F_{r}\equiv\langle \psi|\rho_{0}\left(\tau\right)|\psi\rangle
=\sum_{n,n'=0}^{d-1}\sum_{r=0}^{\text{min}\{n,n'\}}\alpha_{n}\alpha_{n'-r}\alpha_{n'}^{*}\alpha_{n-r}^{*}f_{0}\left(r,n\right)f_{0}^{*}\left(r,n'\right).
\end{equation}
In order to quantify information transport more precisely, we need to calculate the average transmission and reflection fidelities over all initial pure states. Characterizing a $d$-dimensional pure state by means of the Hurwitz-parametrization method \cite{Hur1,Hur2}, then yields
%%%%%%%%%%%%%%%%%%%%%%%%%%%%%%%%%%%%%%%%%%%%%%%%%%%%%%%%%%%%%%%%%%%%%%%%%%%%%%%%%%%%%%%%%%%
\begin{figure}[!ht]%[tpb]
\begin{center}
\includegraphics[width=8.5cm,angle=0]{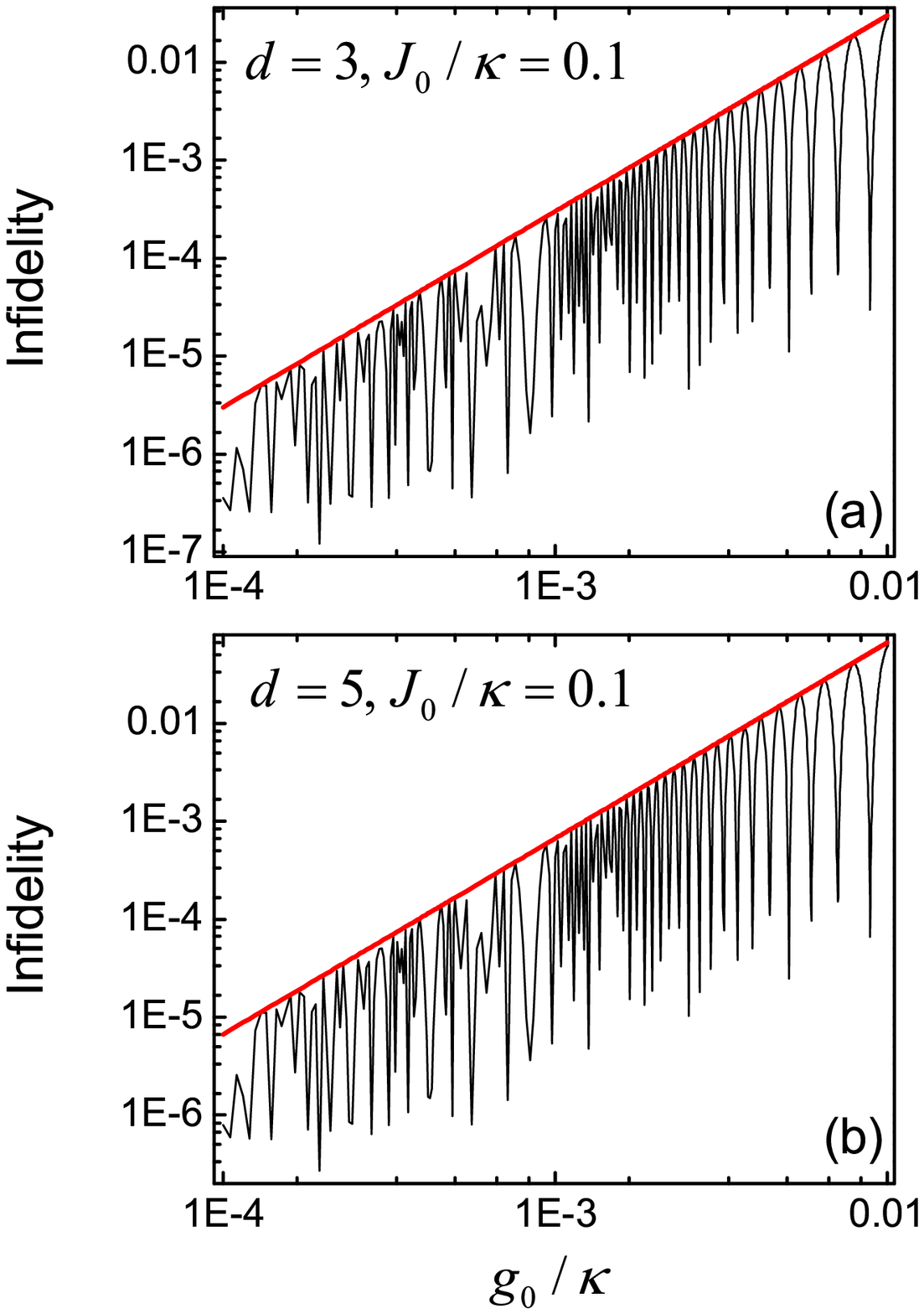}
\caption{(color online) Numerical results of the average reflection infidelity, $\langle \sigma_{r}\rangle=1-\langle F_{r}\rangle$, as a function of $g_{0}/\kappa$ with either (a) $d=3$ or (b) $d=5$ in the coupled case of $J_{0}/\kappa=0.1$. The analytic upper bounds are represented by the bold red lines. Here, the evolution time $t=\tau$ and we choose $N=7$, $m=3$.}\label{figa3}
\end{center}
\end{figure}
%%%%%%%%%%%%%%%%%%%%%%%%%%%%%%%%%%%%%%%%%%%%%%%%%%%%%%%%%%%%%%%%%%%%%%%%%%%%%%%%%%%%%%%%%%%%%%
\begin{equation}\label{eq:ft}
\langle F_{t}\rangle=\frac{1}{d\left(d+1\right)}\left[\sum_{n=0}^{d-1}\sum_{r=0}^{n}|f_{N+1}\left(r,n\right)|^{2}+\sum_{n,n'=0}^{d-1}\left(-1\right)^{\left(n+n'\right)z}
f_{N+1}\left(0,n\right)f_{N+1}^{*}\left(0,n'\right)\right]
\end{equation}
and
\begin{equation}
\langle F_{r}\rangle=\frac{1}{d\left(d+1\right)}\left[\sum_{n=0}^{d-1}\sum_{r=0}^{n}|f_{0}\left(r,n\right)|^{2}
+\sum_{n,n'=0}^{d-1}f_{0}\left(0,n\right)f_{0}^{*}\left(0,n'\right)\right].
\end{equation}
In Fig. \ref{figa1} we plot the average transmission fidelity and the average reflection fidelity. It is found that they exhibit similar behaviors to what we have observed in Fig. \ref{fig2}, except that, in the limit $J_{0}\ll g_{0}$, the average transmission fidelity is close to $1/d$ instead. This difference arises from the fact that the vacuum component of the superposition state $|\psi\rangle$ has an average population probability of $1/d$ after we average over all initial pure states. Hence, controllable transport can be achieved not only for the Fock states but also for arbitrary multiphoton quantum states.

We now consider the quantum information leakage. Specifically, we use an average transmission infidelity $\langle \sigma_{t}\rangle=1-\langle F_{t}\rangle$ for the uncoupled case, and an average reflection infidelity $\langle \sigma_{r}\rangle=1-\langle F_{r}\rangle$ for the coupled case, to quantify the quantum information leakage. It follows, on combining Eqs. (\ref{eq:fmu}), (\ref{eq:Deltat}) and (\ref{eq:ft}), that the average transmission infidelity $\langle \sigma_{t}\rangle$ is
\begin{equation}
\langle \sigma_{t}\rangle=\frac{2d\left(d-1\right)}{d+1}\Delta_{t},
\end{equation}
thereby having an upper bound,
\begin{equation}
\langle \sigma_{t}\rangle\leq\frac{4d\left(d-1\right)}{d+1}\sum_{k<z}\frac{g_{k}^{2}}{\varepsilon_{k}^{2}}.
\end{equation}
In a similar way, the average reflection infidelity $\langle \sigma_{r}\rangle$ is
\begin{equation}
\langle \sigma_{r}\rangle=\frac{2d\left(d-1\right)}{d+1}\Delta_{r},
\end{equation}
and also has an upper bound
\begin{equation}
\langle \sigma_{r}\rangle\leq\frac{2d\left(d-1\right)}{d+1}\frac{g_{z}^{2}}{J_{z}^{2}}.
\end{equation}

As described in the main text, we compare numerical results of the average transmission and reflection infidelities to their analytical upper bounds as shown in Figs. \ref{figa2} and \ref{figa3}, respectively. We find, as expected, that the quantum information leakage for controllable transport of arbitrary multiphoton states is limited by two upper bounds which can be made arbitrary small, in direct analogy to the cases of Fock states in Figs. \ref{fig3} and \ref{fig4}.

%%%%%%%%%%%%%%%%%%%%%%%%%%%%%%%%%%%%%%%%%%%%%%%%%%%%%%%%%%%%%%%%%%%%%%%%%%%%%%%%%%%%%%%%%%%%%%%%%%%%%%%%%%%%%%%%%%%%%%%%%%%%%%%%%%%%%%%%%%%%%%%%%%%%%%%%%%

%%%%%%%%%%%%%%%%%%%%%%%%%%%%%%%%%%%%%%%%%%%%%%%%%%%%%%%%%%%%%%%%%%%%%%%%%%%%%%%%%%%%%%%%%%%%%%%%%%%%%%%%%%%%%%%%%%%%%%%%%%%%%%%%%%%%%%%%%%%%%%%%%%%%%%%%%%%%%%%%%%%%%%%%%

\begin{thebibliography}{999}
\bibitem{photon1} Bouwmeester, D.; Pan, J. W.; Mattle, K.; Eibl, M.; Weinfurter, H.; Zeilinger, A. Experimental quantum teleportation. {\it Nature} (London) {\bf 1997}, 390, 575-579.
\bibitem{photon2} Naik, D. S.; Peterson, C. G.; White, A. G.; Berglund, A. J.; Kwiat, P. G. Entangled State Quantum Cryptography: Eavesdropping on the Ekert Protocol. {\it Phys. Rev. Lett.} {\bf 2000}, 84, 4733-4736.
\bibitem{photon3} Tittel, W.; Brendel, J.; Zbinden, H.; Gisin, N. Quantum Cryptography Using Entangled Photons in Energy-Time Bell States. {\it Phys. Rev. Lett.} {\bf 2000}, 84, 4737-4740.
\bibitem{photon4} Duan, L. M.; Lukin, M. D.; Cirac, J. I.; Zoller, P. Long-distance quantum communication with atomic ensembles and linear optics. {\it Nature} (London) {\bf 2001}, 414, 413-418.
\bibitem{photon5} Deng, F. G.; Long, G. L.; Liu, X. S. Two-step quantum direct communication protocol using the Einstein-Podolsky-Rosen pair block. {\it Phys. Rev. A} {\bf 2003}, 68, 042317.
\bibitem{photon6} Salemian S.; Mohammadnejad, S. An error-free protocol for quantum entanglement distribution in long-distance quantum communication. {\it Chin. Sci. Bull.} {\bf 2011}, 56, 618-625.
\bibitem{photon7} Zheng C.; Long, G. F. Quantum secure direct dialogue using Einstein-Podolsky-Rosen pairs. {\it Sci. China-Phys. Mech. Astron.} {\bf 2014}, 57, 1238-1243.
\bibitem{nuclear1} Zhang, J. F.; Long, G. L.; Zhang, W.; Deng, Z. W.; Liu, W. Z.; Lu, Z. H. Simulation of Heisenberg XY interactions and realization of a perfect state transfer in spin chains using liquid nuclear magnetic resonance. {\it Phys. Rev. A} {\bf 2005}, 72, 012331.
\bibitem{nuclear2} Feng, G. R.; Xu, G. F.; Long, G. L. Experimental Realization of Nonadiabatic Holonomic Quantum Computation. {\it Phys. Rev. Lett.} {\bf 2013}, 110, 190501.
\bibitem{NV1} Childress, L.; Dutt, M. V. G.; Taylor, J. M.; Zibrov, A. S.; Jelezko, F.; Wrachtrup, J.; Hemmer, P. R.; Lukin, M. D. Coherent Dynamics of Coupled Electron and Nuclear Spin Qubits in Diamond. {\it Science} {\bf 2006}, 314, 281-285.
\bibitem{NV2} Zagoskin, A. M.; Johansson, J. R.; Ashhab, S.; Nori, F. Quantum information processing using frequency control of impurity spins in diamond. {\it Phys. Rev. B} {\bf 2007}, 76, 014122.
\bibitem{NV3} Bermudez, A.; Jelezko, F.; Plenio, M. B.; Retzker, A. Electron-Mediated Nuclear-Spin Interactions between Distant Nitrogen-Vacancy Centers. {\it Phys. Rev. Lett.} {\bf 2011}, 107, 150503.
\bibitem{NV4} Yao, N. Y.; Jiang, L.; Gorshkov, A. V.; Maurer, P. C.; Giedke, G.; Cirac, J. I.; Lukin, M. D. Scalable architecture for a room temperature solid-state quantum information processor. {\it Nat. Commun.} {\bf 2012}, 3, 800.
\bibitem{NV5} Wang, P. F.; Ju, C. Y.; Shi, F. Z.; Du, J. F. Optimizing ultrasensitive single electron magnetometer based on nitrogen-vacancy center in diamond. {\it Chin. Sci. Bull.} {\bf 2013}, 58 2920-2923.
\bibitem{NV6} Doherty, M. W.; Manson, N. B.; Delaney, P.; Jelezko, F.; Wrachtrup, J.; Hollenberg, L. C. L. The nitrogen-vacancy colour centre in diamond. {\it Phys. Rep.} {\bf 2013} 528, 1-45.
\bibitem{flux1} Liu, Y. X.; Wei, L. F.; Tsai, J. S.; Nori, F. Controllable Coupling between Flux Qubits. {\it Phys. Rev. Lett.} {\bf 2006}, 96, 067003.
\bibitem{flux2} Niskanen, A. O.; Harrabi, K.; Yoshihara, F.; Nakamura, Y.; Lloyd, S.; Tsai, J. S. Quantum Coherent Tunable Coupling of Superconducting Qubits. {\it Science}, {\bf 2007}, 316, 723-726.
\bibitem{flux3} Ashhab, S. {\it et al}. Interqubit coupling mediated by a high-excitation-energy quantum object. {\it Phys. Rev. B} {\bf 2008}, 77, 014510.
\bibitem{flux4} Xiong, W.; Jin, D. Y.; Jing, J.; Lam, C. H.; You, J. Q. Controllable coupling between a nanomechanical resonator and a coplanar-waveguide resonator via a superconducting flux qubit. {\it Phys. Rev. A} {\bf 2015}, 92, 032318.
\bibitem{STLR1} Sillanp\"{a}\"{a}, M. A.; Park, J. I.; Simmonds, R. W. Coherent quantum state storage and transfer between two phase qubits via a resonant cavity. {\it Nature} (London) {\bf 2007}, 449, 438-442.
\bibitem{STLR2} Majer J. {\it et al}. Coupling superconducting qubits via a cavity bus. {\it Nature} (London) {\bf 2007}, 449, 443-447.
\bibitem{STLR3} Zhou, L.; Gong, Z. R.; Liu, Y. X.; Sun, C. P.; and Nori, F. Controllable Scattering of a Single Photon inside a One-Dimensional Resonator Waveguide. {\it Phys. Rev. Lett.} {\bf 2008}, 101, 100501.
\bibitem{STLR4} Nataf, P.; Ciuti, C. Protected Quantum Computation with Multiple Resonators in Ultrastrong Coupling Circuit QED. {\it Phys. Rev. Lett.} {\bf 2011}, 107, 190402.
\bibitem{STLR5} Yang, C. P.; Su, Q. P.; Nori, F. Entanglement generation and quantum information transfer between spatially-separated qubits in different cavities. {\it New J. Phys.} {\bf 2013}, 15, 115003.
\bibitem{TM1} Armani, D. K.; Kippenberg, T. J.; Spillane, S. M.; Vahala, K. J. Ultra-high-Q toroid microcavity on a chip. {\it Natrue} (London) {\bf 2003}, 421, 925-928.
\bibitem{TM2} Peng, B. {\it et al}. Parity-time-symmetric whispering-gallery microcavities. {\it Nat. Phys.} {\bf 2014}, 10, 394-398.
\bibitem{TM3} Peng, B. {\it et al}. Loss-induced suppression and revival of lasing. {\it Science} {\bf 2014}, 346, 328-332.
\bibitem{TM4} Jing, H.; Oezdemir, S. K.; Lu, X. Y.; Zhang, J.; Yang, L.; Nori, F. PT-Symmetric Phonon Laser. {\it Phys. Rev. Lett.} {\bf 2014}, 113, 053604.
\bibitem{PNA} Lee, C.; Tame, M.; Noh, C.; Lim, J.; Maier, S. A.; Lee, J.; Angelakis, D. G. Robust-to-loss entanglement generation using a quantum plasmonic nanoparticle array. {\it New J. Phys.} {\bf 2013}, 15, 083017.
\bibitem{node1} Cirac, J. I.; Zoller, P.; Kimble, H. J.; Mabuchi, H. Quantum State Transfer and Entanglement Distribution among Distant Nodes in a Quantum Network. {\it Phys. Rev. Lett.} {\bf 1997}, 78, 3221-3224.
\bibitem{node2} Serafini, A.; Mancini, S.; Bose, S. Distributed Quantum Computation via Optical Fibers. {\it Phys. Rev. Lett.} {\bf 2006}, 96, 010503.
\bibitem{node3} Kimble, H. J. The quantum internet, {\it Nature} (London) {\bf 2008}, 453, 1023-1030.
\bibitem{MB1} Hartmann, M. J.; Brand\~{a}o, F. G. S. L.; Plenio, M. B. Quantum many-body phenomena in coupled cavity arrays. {\it Laser Photon. Rev.} {\bf 2008},  2, 527-556.
\bibitem{MB2} Hafezi, M.; Demler, E. A.;Lukin, M. D.; Taylor, J. M. Robust optical delay lines with topological protection. {\it Nat. Phys.} {\bf 2011}, 7, 907-912 (2011).
\bibitem{MB3} Georgescu, I. M.; Ashhab, S.; Nori, F. Quantum simulation. {\it Rev. Mod. Phys.} {\bf 2014}, 86, 153-185.
\bibitem{MB4} Douglas, J. S.; Habibian, H.; Hung, C.-L.; Gorshkov, A. V.; Kimble, H. J.; Chang, D. E. Quantum many-body models with cold atoms coupled to photonic crystals. {\it Nat. Photon.} {\bf 2015}, 9, 326-331.
\bibitem{CCA_switch1} Wang, Z. H.; Li, Y.; Zhou, D. L.; Sun, C. P.; Zhang, P. Single-photon scattering on a strongly dressed atom. {\it Phys. Rev. A} {\bf 2012}, 86, 023824.
\bibitem{CCA_switch2} Zhou, L.; Yang, L. P.; Li, Y.; Sun, C. P. Quantum Routing of Single Photons with a Cyclic Three-Level System. {\it Phys. Rev. Lett.} {\bf 2013}, 111, 103604.
\bibitem{CCA_switch3} Lombardo, F.; Ciccarello, F.; Palma, G. M. Photon localization versus population trapping in a coupled-cavity array. {\it Phys. Rev. A} {\bf 2014}, 89, 053826.
\bibitem{CCA_switch4} Hai, L.; Tan, L.; Feng, J. S.; Xu, W. B.; Wang, B. Single photon transport properties in coupled cavity arrays nonlocally coupled to a two-level atom in the presence of dissipation. {\it Chin. Phys. B} {\bf 2014}, 23, 024202.
\bibitem{CCA_switch5} Lu, J.; Wang, Z. H.; Zhou, L. T-shaped single-photon router. {\it Opt. Express} {\bf 2015}, 23, 22955-22962.
\bibitem{single1} Xu, D. Z.; Ian, H.; Shi, T.; Dong, H.; Sun, C. P. Photonic Feshbach resonance. {\it Sci. China-Phys. Mech. Astron.} {\bf 2010}, 53, 1234-1238.
\bibitem{single2} Zhu, W.; Wang, Z. H.; Zhou, D. L. Multimode effects in cavity QED based on a one-dimensional cavity array. {\it Phys. Rev. A} {\bf 2014}, 90, 043828.
\bibitem{single3} Qin, W; Nori, F. Controllable single-photon transport between remote coupled-cavity arrays. {\it Phys. Rev. A} {\bf 2016}, 93, 032337.
\bibitem{multi1_1} Longo, P.; Schmitteckert, P.; Busch, K. Few-Photon Transport in Low-Dimensional Systems: Interaction-Induced Radiation Trapping. {\it Phys. Rev. Lett.} {\bf 2010}, 104, 023602.
\bibitem{multi1_2} Longo, P.; Schmitteckert, P.; Busch, K. Few-photon transport in low-dimensional systems. {\it Phys. Rev. A} {\bf 2011}, 83, 063828.
\bibitem{multi2_1} Roy, D. Few-photon optical diode. {\it Phys. Rev. B} {\bf 2010}, 81, 155117.
\bibitem{multi2_2} Roy, D. Two-Photon Scattering by a Driven Three-Level Emitter in a One-Dimensional Waveguide and Electromagnetically Induced Transparency. {\it Phys. Rev. Lett.} {\bf 2011}, 106, 053601.
\bibitem{multi3} Shi, T.; Chang, D. E.; Cirac, J. I. Multiphoton-scattering theory and generalized master equations. {\it Phys. Rev. A} {\bf 2015}, 92, 053834.
\bibitem{un_qc1} Luo, M. X.; Wang, X. J. Universal quantum computation with qudits. {\it Sci. China-Phys. Mech. Astron.} {\bf 2014}, 57, 1712-1717.
\bibitem{un_qc2} Qin, W.; Wang, C.; Cao, Y.; Long, G. L. Multiphoton quantum communication in quantum networks. {\it Phys. Rev. A} {\bf 2014}, 89, 062314.
\bibitem{un_qc3} Hua, M.; Tao, M. J.; Deng, F. G. Efficient generation of NOON states on two microwave-photon resonators. {\it Chin. Sci. Bull.} {\bf 2014}, 59, 2829-2834.
\bibitem{gfock} Hofheinz, M. {\it et al}. Generation of Fock states in a superconducting quantum circuit. {\it Nature} {\bf 2008}, 454, 310-314.
\bibitem{numcals} In order to plot the transmission and reflection infidelities, we numerically perform the exact diagonalization of the coupling matrix $A$, and then calculate these infidelities according to Eq. (26), Eq. (27), and the definitions of such infidelities. The upper bounds, however, are straightforwardly plotted according to Eqs. (47) and (54). In addition, we plot these quantities at a specific evolution time, $t=\tau\equiv\pi/\sqrt{2}g_{z}$, to make the quantum state of either the transmitted or reflected photons remain unchanged.
\bibitem{ut1} Lieb, E.; Schultz, T.; Mattis, D. Two Soluble Models of an Antiferromagnetic Chain. {\it Ann. Phys.} (NY) {\bf 1961}, 16, 407-466.
\bibitem{ut2} Christandl, M.; Datta, N.; Ekert, A.; Landahl, A. J. Perfect State Transfer in Quantum Spin Networks. {\it Phys. Rev. Lett.} {\bf 2004}, 92, 187902.
\bibitem{spin0} Yao, N. Y.; Jiang, L.; Gorshkov, A.V.; Gong, Z.-X.; Zhai, A.; Duan, L.-M.; Lukin, M. D. Robust Quantum State Transfer in Random Unpolarized Spin Chains. {\it Phys. Rev. Lett.} {\bf 2011}, 106, 040505.
\bibitem{spin1} Bose, S. Quantum Communication through an Unmodulated Spin Chain. {\it Phys. Rev. Lett.} {\bf 2003}, 91, 207901.
\bibitem{spin2} Liu, Y.; Zhou, D. L. Optimized quantum state transfer through an XY spin chain.{\it Phys. Rev. A} {\bf 2014}, 89, 062331.
\bibitem{spin3} Qin, W.; Li, J. L.; Long, G. L. High-dimensional quantum state transfer in a noisy network environment. {\it Chin. Phys. B} {\bf 2015}, 24, 040305.
\bibitem{spin4} Qin, W.; Wang, C.; Zhang, X. D. Protected quantum-state transfer in decoherence-free subspaces. {\it Phys. Rev. A.} {\bf 2015}, 91, 042303.
\bibitem{spin5} Fukuhara, T.; Hild, S.; Zeiher, J.; Schau{\ss}, P.; Bloch, I.; Endres, M.; Gross, C. Spatially Resolved Detection of a Spin-Entanglement Wave in a Bose-Hubbard Chain. {\it Phys. Rev. Lett.} {\bf 2015}, 115, 035302.
\bibitem{Hur1} \.{Z}yczkowski, K.; Sommers, H. Induced measures in the space of mixed quantum states. {\it J. Phys. A: Math. Gen.} {\bf 2001}, 34, 7111-7125.
\bibitem{Hur2} Yang, Z.; Gao, M.; Qin, W. Transfer of high-dimensional quantum state through an XXZ-Heisenberg quantum spin chain. {\it Int. J. Mod. Phys. B} {\bf 2015}, 29, 1550207.
\end{thebibliography}
\end{document}